# Erratum: Spectral representation for the effective macroscopic response of a polycrystal: application to third-order non-linear susceptibility [J. Phys. Cond. Matt. 26, 10323-10334 (1999)]

S.V. Barabash[1;2] and D. Stroud[1]

[1] *Department of Physics, The Ohio State University, Columbus, OH 43210*

[2] *Present address: Intermolecular, Inc., 3011 N. 1st St., San Jose, CA 95134*

E-mail: stroud@physics.osu.edu and sergey.barabash@intermolecular.com

(Dated: August 10, 2015)

In our paper[1], we show that the spectral representation for isotropic two-component composites also applies to uniaxial polycrystals. We have learned that this result was, in fact, first conjectured by G.W. Milton in Appendix E of Ref. 2 below. While our derivation is more detailed, our result for the spectral function is the same as Milton's. We very much regret not having been aware of this work at the time of writing our paper (Ref.1).

[1] S. Barabash and D. Stroud, J. Phys. Cond. Matt. **26**, 10323-10334 (1999); arXiv:cond-mat/9910246v1.
[2] G.W. Milton, J. Appl. Phys. **52**, 5294 (1981).

# Spectral Representation for the Effective Macroscopic Response of a Polycrystal: Application to Third-Order Nonlinear Susceptibility

S. Barabash and D. Stroud

*Department of Physics, The Ohio State University, Columbus, Ohio 43210*

(July 29, 2013)

## Abstract

We extend the spectral theory used for the calculation of the effective linear response functions of composites to the case of a polycrystalline material with uniaxially anisotropic microscopic symmetry. As an application, we combine these results with a nonlinear decoupling approximation as modified by Ma *et al*, to calculate the third-order nonlinear optical susceptibility of a uniaxial polycrystal, assuming that the effective dielectric function of the polycrystal can be calculated within the effective-medium approximation.

## I. INTRODUCTION

Almost twenty years ago, Bergman [1] developed the spectral approach for calculating the dielectric constant and other linear response functions of a two-component composite. His approach was to study the analytical properties of the effective dielectric constant, viewed as a function of the ratio of the dielectric constants of the constituents. Among other results, he showed that all poles of this function can be expressed as eigenvalues of a certain linear boundary-value problem, while the residues of those poles are given as certain integrals over the corresponding eigenfunctions. Bergman's theory actually describes a wide class of mathematically similar physical problems in which a divergence-free field appears as a linear response to the gradient of a potential; thus, it can be used to find the effective electrical and thermal conductivities, magnetic permeability and many other effective parameters described by mathematically equivalent equations.

In some microgeometries, this eigenproblem can be solved by expanding the exact eigenfunctions of the composite in terms of the individual grain eigenfunctions. This approach has been used to calculate the effective parameters of granular composites corresponding to several different ordered microgeometries [1]. However, the spectral representation is often useful even in composites where the microgeometry is not known exactly; in such cases, one must generally resort to various approximations in order to calculate the relevant spectral functions. Moreover, the spectral approach is not limited to problems involving *linear* response. For example, some of the effective macroscopic nonlinear response functions can be expressed in terms of linear response functions of the composite, and certain geometric factors [5]. Recently, this connection has been employed [3], together with certain approximations for linear composites, to use the spectral representation in calculating these nonlinear response functions.

The present paper is directed towards the linear and nonlinear dielectric response of *polycrystalline* materials. Such materials are not generally thought of as composite media, but in fact they behave like composites. The reason is that even though each crystallite is made of the same material, it has a different spatial orientation and hence has different constitutive properties referred to axes fixed in the lab coordinate system. In particular, in this paper we extend the spectral theory to describe both the linear and nonlinear response of a polycrystalline material.

By a polycrystal, we mean a material with anisotropic transport properties, such that the crystal symmetry axes vary in direction from point to point in space. Several previous workers (see, for example, [6]) have described polycrystals as composite materials. In the present work, we further restrict our discussion to polycrystals of uniaxial materials. In this case, two of the three principal components of the dielectric tensor are equal, and it is more straightforward to develop a spectral representation for the effective properties. This restriction to uniaxial materials still leave many classes of crystalline materials open to study. In particular, the theory should satisfactorily describe such classes of materials as the quasi-one-dimensional organic conductors [6], or the quasi-planar or $Cu_2O$ -based high-$T_c$ superconductors.

We will use this approach not only to describe the linear properties, but also to calculate the enhancement of the third-order nonlinear susceptibility of a polycrystal. Although this enhancement has been previously discussed theoretically [4], the treatment presented in that previous discussion needs to be modified in the case of a complex-valued susceptibility, as has been pointed out by Ma *et. al.* [3].

The remainder of this paper is organized as follows. Section II describes the extension of the spectral theory to uniaxial polycrystalline materials. The application of this theory to the nonlinear response of polycrystals is given in Section III, followed by a numerical example in Section IV.

## II. SPECTRAL THEORY FOR THE EFFECTIVE MACROSCOPIC LINEAR RESPONSE OF A POLYCRYSTAL

We consider a polycrystalline dielectric material characterized by a position-dependent uniaxially symmetric dielectric tensor, which we express in the form

$$\overleftrightarrow{\epsilon}(\mathbf{r}) = \overleftrightarrow{R}^{-1}(\mathbf{r})\overleftrightarrow{\epsilon}_d\overleftrightarrow{R}(\mathbf{r}), \tag{1}$$

where

$$\overleftrightarrow{\epsilon}_d = \begin{pmatrix} \epsilon_1 & 0 & 0 \\ 0 & \epsilon_2 & 0 \\ 0 & 0 & \epsilon_2 \end{pmatrix} \tag{2}$$

is the dielectric tensor in the frame of principal axis and $\overleftrightarrow{R}(\mathbf{r})$ is a position-dependent orthogonal matrix characterizing the microstructure of a particular specimen (specifically, it describes the local orientation of the principal axes with respect to the lab axes). If the sample is macroscopically isotropic, it is reasonable to assume that on a large scale its dielectric behavior can be characterized by a scalar dielectric constant $\epsilon_e$. $\epsilon_e$ may be defined by the relation

$$\mathbf{D_0} \equiv \frac{1}{V} \int \mathbf{D}(\mathbf{x}) d^3 x = \epsilon_e \mathbf{E_0}. \tag{3}$$

where $\mathbf{E_0} \equiv \frac{1}{V} \int \mathbf{E}(\mathbf{x}) d^3 x$ is the space-averaged electric field. We assume that $\mathbf{E_0}$ is real and is directed along the $z$ axis: $\mathbf{E_0} = E_0 \hat{z}$. In general, the fields $\mathbf{E}$ and $\mathbf{D}$ (as well as the dielectric tensor) are represented as complex quantities; the physical fields are related to them through $\mathbf{E_{phys}}(\mathbf{x}) = Re\left(\mathbf{E}(\mathbf{x})e^{-i\omega t}\right)$, $\mathbf{D_{phys}}(\mathbf{x}) = Re\left(\mathbf{D}(\mathbf{x})e^{-i\omega t}\right)$.

In the quasi-static approximation, the electric field is given by the negative gradient of a scalar potential. We may express this potential through the relation by $\mathbf{E} = -E_0 \nabla \phi$, where $E_0$ is real and $\phi$ is the solution to the following boundary-value problem:

$$\begin{aligned} \nabla \cdot (\overleftrightarrow{\epsilon}(\mathbf{x}) \nabla \phi) &= 0 \text{ in } V, \\ \phi &= \phi_0 \equiv -z \text{ on } S, \end{aligned} \tag{4}$$

where $S$ is the boundary surrounding $V$. Using the boundary conditions for $\phi$ and the Maxwell's equation $\nabla \cdot \mathbf{D} = 0$, we can show that [5]:

$$\begin{aligned} \frac{1}{V} \int \mathbf{D} \cdot \mathbf{E} dV &= \frac{1}{V} \int \mathbf{D} \cdot (-E_0 \nabla \phi) dV = \frac{1}{V} \int \nabla \cdot [\mathbf{D}(-E_0 \phi)] dV \\ &= \frac{1}{V} \oint \mathbf{D}(-E_0 \phi) \cdot \mathbf{dS} = \frac{1}{V} \oint \mathbf{D}(-E_0 \phi_0) \cdot \mathbf{dS} \\ &= \frac{1}{V} \int \nabla \cdot [\mathbf{D}(-E_0 \phi_0)] dV = \frac{1}{V} \int \mathbf{D} \cdot (E_0 \nabla z) dV = \mathbf{D_0} \cdot \mathbf{E_0}. \end{aligned} \tag{5}$$

Therefore, the definition (3) is equivalent to

$$\epsilon_e = \frac{1}{V} \int \frac{\mathbf{E} \cdot \mathbf{D}}{E_0^2} dV. \tag{6}$$

The result (6) is the equation we use below to express $\epsilon_e$ in terms of eigenvalues of a linear operator.

In order to achieve this reduction, we first note that instead of the position-dependent tensor $\overleftrightarrow{\epsilon}$ we can use

$$\frac{\overleftrightarrow{\epsilon}}{\epsilon_2} = \overleftrightarrow{1} - u \overleftrightarrow{R}^{-1} \overleftrightarrow{C} \overleftrightarrow{R}, \tag{7}$$

where the parameter $u$ is defined by

$$u \equiv 1 - \frac{\epsilon_1}{\epsilon_2}, \tag{8}$$

and $\overleftrightarrow{C}$ is a matrix with $C_{11} = 1$ and other components equal to zero. We can use this result to rewrite the first line of (4) as

$$\nabla^2 \phi = u \left(\nabla \overleftrightarrow{(R^{-1})}\right)_1 \left(\overleftrightarrow{R} \nabla\right)_1 \phi \equiv u \partial_i R_{1i} R_{1j} \partial_j \phi, \tag{9}$$

where we have used $\left(\overleftrightarrow{R}^{-1}\right)_{ij} = \left(\overleftrightarrow{R}\right)_{ji}$, and also employed the convention that repeated indices are summed over. From eq. (9), we see that (4) is equivalent to the integral equation

$$\phi = -z + u\Gamma\phi. \tag{10}$$

Here the linear operator $\Gamma$ is defined by its effect on a function $\phi$ through

$$\Gamma\phi \equiv -\int d^3r' G(\mathbf{r},\mathbf{r}') \left(\nabla'\overleftrightarrow{R}^{-1}(\mathbf{r}')\right)_1 \left(\overleftrightarrow{R}(\mathbf{r}')\nabla'\phi(\mathbf{r}')\right)_1 \tag{11}$$

and $G(\mathbf{r},\mathbf{r}')$ is a Green's function for the Laplace operator:

$$\begin{aligned} \nabla^2 G(\mathbf{r},\mathbf{r}') &= -\delta^3(\mathbf{r}-\mathbf{r}') \text{ for } \mathbf{r} \text{ in } V \\ G &= 0 \text{ for } \mathbf{r} \text{ on the boundary.} \end{aligned} \tag{12}$$

It is now convenient to define a scalar product of two functions by

$$\langle\phi|\psi\rangle = \int dV \left(\nabla\phi^*\overleftrightarrow{R}^{-1}\right)_1 \left(\overleftrightarrow{R}\nabla\psi\right)_1, \tag{13}$$

With this definition, we can show that $\Gamma$ is self-adjoint, non-negative, bounded linear operator. To show the self-adjoint property, we integrate (11) by parts using the boundary conditions for $G$ to obtain

$$\Gamma\phi = \int dV' \left(\partial_i' G(\mathbf{r},\mathbf{r}')\right) R_{1i}(\mathbf{r}') R_{1j}(\mathbf{r}') \partial_j' \phi(\mathbf{r}'). \tag{14}$$

Then, using the fact that $G$ satisfies $G(\mathbf{r},\mathbf{r}') = G(\mathbf{r}',\mathbf{r})$, we find that

$$\begin{aligned} \langle\phi|\Gamma\psi\rangle &= \int dV \int dV' \partial_i \phi(\mathbf{r}) R_{1i} R_{1j} \partial_j \left(\partial_k' G(\mathbf{r},\mathbf{r}') R'_{1k} R'_{1l} \partial_l' \psi(\mathbf{r}')\right) \\ &= \int dV \int dV' R_{1i} R_{1j} R'_{1k} R'_{1l} \partial_j \left(\partial_k' G(\mathbf{r},\mathbf{r}')\right) \partial_i \phi(\mathbf{r}) \partial_l' \psi(\mathbf{r}') \\ &= \langle\Gamma\phi|\psi\rangle. \end{aligned} \tag{15}$$

To prove that $\Gamma$ is real, bounded and non-negative, we consider the eigenvalue problem

$$\begin{aligned} \Gamma\phi_i(\mathbf{r}) &= s_i\phi_i(\mathbf{r}) \text{ for } \mathbf{r} \text{ in } V, \\ \phi_i &= 0 \text{ for } \mathbf{r} \text{ on the boundary,} \end{aligned} \tag{16}$$

where $s_i \equiv 1/u_i$ and $u_i$ is the value of $u$ at one of the eigenstates $\phi_i$ (the so-called "electrostatic resonances"). The physical significance of the latter has been discussed elsewhere [1]. Next, we note that the problem defined by eq. (16) is equivalent to the problem

$$\begin{aligned} \nabla \cdot \left[\left(s_i\overleftrightarrow{1} - \overleftrightarrow{R}^{-1}\overleftrightarrow{C}\overleftrightarrow{R}\right)\nabla\phi_i\right] &= 0, \\ \phi_i &= 0 \text{ at the boundary.} \end{aligned} \tag{17}$$

as can be seen by comparing the steps going from (4) to (10). But from eq. (17), we can write

$$\begin{aligned} 0 &= \int dV \phi_i^* \nabla \cdot \left(s_i\overleftrightarrow{1} - \overleftrightarrow{R}^{-1}\overleftrightarrow{C}\overleftrightarrow{R}\right)\nabla\phi_i \\ &= -\int dV \left(s_i|\nabla\phi_i|^2 - \left|\left(\overleftrightarrow{R}\nabla\phi_i\right)_1\right|^2\right), \end{aligned} \tag{18}$$

from which it follows that $0 \leq s_i < 1$. The limiting case $s_i = 1$ ($\epsilon_1 = 0$) could be realized only if the tensor $\overleftrightarrow{R}$ were position-independent. But this would lead to $\epsilon_e \equiv \overleftrightarrow{\epsilon}(\mathbf{r})$, i. e., a tensor, which contradicts our assumption that $\epsilon_e$ is a scalar value. The fact that the eigenvalues of (16) are limited to the semiclosed segment $[0, 1)$ proves our statement that $\Gamma$ is real, bounded, and non-negative.

From the properties of $\Gamma$, we conclude that the eigenfunctions $|\phi_i\rangle$ of (16) form a complete orthogonal set with respect to the scalar product (13). Hence, the solution to eq. (10) can be expressed (except on the boundary) as

$$|\phi\rangle = (u\Gamma - 1)^{-1} |z\rangle \equiv \sum_i \left(\frac{s}{s_i - s}\right) \frac{|\phi_i\rangle \langle \phi_i|z\rangle}{\langle \phi_i|\phi_i\rangle}. \tag{19}$$

We can now use this equation to find an analytical representation for $\epsilon_e$.

We begin by using Green's theorem, the boundary conditions in (4) for $\phi$, and the Maxwell equation $\nabla \cdot \mathbf{D} = 0$ to rewrite eq. (6). The result is

$$\begin{aligned}
\frac{\epsilon_e}{\epsilon_2} &= \frac{1}{\epsilon_2 V E_0^2} \int (-E_0 \nabla \phi) \cdot \mathbf{D} dV = \frac{1}{\epsilon_2 V E_0} \oint z \mathbf{D} \cdot d\mathbf{S} = \frac{1}{\epsilon_2 V E_0} \int \hat{\mathbf{z}} \cdot \mathbf{D} dV \\
&= \frac{-1}{V} \int \hat{\mathbf{z}} \cdot \left(\left(\overleftrightarrow{1} - u\overleftrightarrow{R}^{-1}\overleftrightarrow{C}\overleftrightarrow{R}\right) \nabla \phi\right) dV = \frac{1}{V} \oint z\hat{\mathbf{z}} \cdot d\mathbf{S} + \frac{u}{V} \int \left(\nabla z \overleftrightarrow{R}^{-1}\right)_1 \left(\overleftrightarrow{R} \nabla \phi\right)_1 dV \quad (20) \\
&= 1 + \frac{u}{V} \langle z|\phi\rangle.
\end{aligned}$$

If we now introduce a function

$$F(s) = 1 - \frac{\epsilon_e}{\epsilon_2}, \tag{21}$$

then on substituting (19) we find

$$\begin{aligned}
F(s) &= -\frac{u}{V}\langle z|\phi\rangle = \frac{1}{V}\langle z|\frac{1}{s-\Gamma}|z\rangle \\
&= \frac{1}{V} \sum_i \frac{|\langle z|\phi_i\rangle|^2}{\langle \phi_i|\phi_i\rangle} \left(\frac{1}{s - s_i}\right).
\end{aligned} \tag{22}$$

This final result is identical in form to Bergman's expression for the analogous function in scalar composite materials. The only difference lies in the definition of the scalar product (13).

### III. APPLICATION TO THIRD ORDER NONLINEAR RESPONSE OF POLYCRYSTALS

As has been suggested by several authors (see, for example, [7]), the nonlinear susceptibilities of composite materials may be hugely enhanced by large fluctuations in the local electric field in these materials. The basic idea is as follows: since these nonlinear susceptibilities depend on higher powers of the local electric field than does the linear dielectric function $\epsilon_e$, any enhancement of that field will produce an even larger enhancement in those susceptibilities than in $\epsilon_e$.

A theoretical description of this enhancement has been given by several authors, initially for isotropic composite materials [5], and more recently for polycrystals [4,6]. The original exact expression given in [5] is generally difficult to evaluate without approximations. One useful approximation involves a decoupling assumption: a certain average of the fourth power of the electric field, which enters the exact expression, is approximated as a product of averages of second powers [8]. However, this decoupling approximation (as well as the original exact expression) must be modified slightly when the material of interest has a complex-valued dielectric tensor. The need for such a modification was first noted by Ma *et al*, who also generalize the approach of [5] for the case of components with complex scalar dielectric functions. In what follows, we further generalize the approach of Ma *et al* [3] to the case of polycrystals, using the results of Sec. II.

We consider a polycrystalline material in which $\mathbf{D}(\mathbf{x}, \omega)$ and $\mathbf{E}(\mathbf{x}, \omega)$ are related by

$$D_i = \epsilon_{ij} E_j + \chi_{ijkl} E_j E_k E_l^*, \tag{23}$$

where we suppress the frequency and position dependence of all quantities and sum over repeated indices. Next, we assume that a sufficiently large sample of this polycrystal can be treated as macroscopically isotropic. Thus, the effective response at the fundamental frequency $\omega$ is given by

$$\mathbf{D_0} = \langle \mathbf{D} \rangle = \epsilon_e \mathbf{E_0} + \chi \left( \mathbf{E_0} \cdot \mathbf{E_0^*} \right) \mathbf{E_0} + \tilde{\chi} \left( \mathbf{E_0} \cdot \mathbf{E_0} \right) \mathbf{E_0^*}, \tag{24}$$

where $\mathbf{E}_0$ is the spatial average of the electric field. In component notation, this may be written

$$D_{0,i} = \epsilon_e E_{0,i} + \chi \delta_{ij} \delta_{kl} E_{0,j}^* E_{0,k} E_{0,l} + \tilde{\chi} \delta_{ij} \delta_{kl} E_{0,j} E_{0,k} E_{0,l}^*, \tag{25}$$

where $\langle ... \rangle$ denotes a volume average and $D_{0,i}$ and $E_{0,i}$ are the $i^{th}$ components of $\mathbf{D}_0$ and $\mathbf{E}_0$. The method of ref. [4] does not permit the two effective susceptibilities to be easily calculated independently, but their sum is readily computed. Generalizing eq. (13) of ref. 4 to the case of finite frequencies, we obtain

$$\chi_e \equiv \chi + \tilde{\chi} = \frac{\langle \chi_{ijkl} E_i E_j E_k E_l^* \rangle}{E_0^4}. \tag{26}$$

where $E_i \equiv E_i(\mathbf{x}, \omega)$ denotes the Cartesian component of the local electric field at frequency $\omega$ in the corresponding *linear* polycrystal.

In this paper we will assume that the fourth-rank tensor $\chi_{ijkl}(\mathbf{x}, \omega)$ has certain symmetry properties which cause many of its components to vanish. Specifically, we will assume that the only non-vanishing components of $\chi_{ijkl}(\mathbf{x}, \omega)$ (in a frame of reference where the coordinate axes are parallel to the local symmetry axes of the crystallite) are those such that the indices are equal in pairs. Then Eq. (26) takes the form

$$\chi_e = \chi_{iijj} \frac{\langle E_i^2 |E_j|^2 \rangle}{E_0^4}, \tag{27}$$

where $E_i = E_i(\mathbf{x})$ is the field component parallel to the i$^{th}$ principal axis at $\mathbf{x}$ (where we suppress the frequency index $\omega$).

We approximate the right-hand side using the nonlinear decoupling approximation (NDA) [4,8], which is specified by the assumption

$$\left\langle E_i^2 |E_j|^2 \right\rangle \approx \left\langle E_i^2 \right\rangle \left\langle |E_j|^2 \right\rangle. \tag{28}$$

Then using the expression $\epsilon_e = \frac{1}{VE_0^2} \sum_{i=1}^{3} \epsilon_i \int E_i(\mathbf{x})^2 dV$ we immediately get [4]:

$$\left\langle E_i^2 \right\rangle = \frac{1}{E_0^2} \left( \frac{\partial \epsilon_e}{\partial \epsilon_i} \right). \tag{29}$$

Here the partial derivative denotes $\partial \epsilon_e(\epsilon_1, \epsilon_2, \epsilon_3)/\partial \epsilon_i$. In the case of a uniaxial material, one should calculate this derivative first with $\epsilon_2 \neq \epsilon_3$, and only then take the limit $\epsilon_2 = \epsilon_3$.

The second average on the right-hand side of (28) can be evaluated [3] with the help of the spectral approach developed in the previous section. From (19) and our definition (13) of the scalar product, we find

$$\langle |E_1|^2 \rangle = \frac{E_0^2}{V} \langle \phi | \phi \rangle = \frac{E_0^2}{V} \sum_j \sum_i \frac{|s|^2}{(s_j - s^*)(s_i - s)} \frac{\langle z | \phi_j \rangle \langle \phi_i | z \rangle}{\langle \phi_j | \phi_j \rangle \langle \phi_i | \phi_i \rangle} \langle \phi_j | \phi_i \rangle$$

$$= \frac{E_0^2}{V} \sum_i \frac{|s|^2}{|s_i - s|^2} \frac{|\langle \phi_i | z \rangle|^2}{\langle \phi_i | \phi_i \rangle}, \tag{30}$$

where we have used the orthogonality condition $\langle \phi_i | \phi_j \rangle = \delta_{ij}$.

To evaluate $\langle |E_2|^2 \rangle$, we note that the boundary conditions (4) for $\phi$ and for $\phi^*$ are the same, since $\phi_0$ is real on the boundary. Therefore, we can use $E^*$ in place of $E$ in the transformations (5), so that the definition (6) becomes

$$\epsilon_e E_0^2 = \frac{1}{V} \int \mathbf{E}^* \cdot \mathbf{D} dV = \epsilon_1 \langle |E_1|^2 \rangle + 2\epsilon_2 \langle |E_2|^2 \rangle, \tag{31}$$

where we used $\langle |E_2|^2 \rangle = \langle |E_3|^2 \rangle$. Hence,

$$\langle |E_2|^2 \rangle = \frac{1}{2} \left( \frac{\epsilon_e}{\epsilon_2} - \frac{\epsilon_1}{\epsilon_2} \frac{\langle |E_1|^2 \rangle}{E_0^2} \right) E_0^2$$

$$= \frac{1}{2} \left( 1 - F(s) - (1 - 1/s) \frac{\langle |E_1|^2 \rangle}{E_0^2} \right) E_0^2 \tag{32}$$

$$= \frac{1}{2} \left( 1 - \sum_i \frac{(|s|^2 - s_i)}{|s - s_i|^2} \frac{|\langle \phi_i | z \rangle|^2}{\langle \phi_i | \phi_i \rangle} \right) E_0^2.$$

Given an approximation for the effective *linear* response function $\epsilon_e$, the above formulas allow us to calculate the enhancement of the non-linear susceptibility.

The simplest approximation for $\epsilon_e(\epsilon_1, \epsilon_2, \epsilon_3)$ is the effective-medium approximation (EMA) [9], which gives

$$\sum_{i=1}^{3} \frac{\epsilon_i - \epsilon_e}{\epsilon_i + 2\epsilon_e} = 0, \tag{33}$$

or for a uniaxial material,

$$\frac{\epsilon_1 - \epsilon_e}{\epsilon_1 + 2\epsilon_e} + 2\frac{\epsilon_2 - \epsilon_e}{\epsilon_2 + 2\epsilon_e} = 0. \tag{34}$$

Then the function $F(s)$ [eq. (21)] is given by

$$F(s) = \frac{3}{4}\left(1 - \sqrt{\frac{s - 8/9}{s}}\right). \tag{35}$$

The corresponding $\Gamma$ operator has a continuous spectrum, so that the sums in (22), (30) and (32) should be replaced by integrals. From

$$F(s) = \int_0^1 \frac{\mu(x)}{s - x} dx \tag{36}$$

we find that

$$\mu(x) = -\frac{1}{\pi} Im[F(x + i0)] = \frac{3}{4\pi}\sqrt{\frac{(x - 8/9)}{x}} \theta(-x)\theta(x - 8/9), \tag{37}$$

where $\theta(x)$ is the usual step function, i. e., $\theta(x) = 1$ for $x > 0$ and $\theta(x) = 0$ for $x \leq 0$. In order to evaluate the effective nonlinear response, this expression should be substituted into the integrals

$$\langle |E_1|^2 \rangle = \int_0^1 dx \frac{|s|^2 \mu(x)}{|s - x|^2} E_0^2, \tag{38}$$

$$\langle |E_2|^2 \rangle = \frac{1}{2}\left(1 - \int_0^1 dx \frac{(|s|^2 - x)\mu(x)}{|s - x|^2}\right) E_0^2. \tag{39}$$

One immediate consequence of the EMA is that the integral in (39) diverges as $s \to 0$ (i. e., as $\epsilon_1/\epsilon_2 \to \infty$), and hence, $\langle |E_2|^2 \rangle$ also diverges in the same limit. This divergence is related to the divergence of $\langle E_2^2 \rangle$ [eq. (29)]; the physical origin of that latter divergence was discussed in [4]). Thus, in the NDA/EMA approximation, if $\chi_{2222} \neq 0$ $\chi_e$ becomes *arbitrarily large* when $\epsilon_1/\epsilon_2 \to \infty$, both at zero and at finite frequencies.

## IV. NUMERICAL EXAMPLE

To illustrate this discussion, we consider a simple model for a polycrystalline quasi-1D conductor. In the high-conductivity direction, we assume a Drude metal with dielectric function

$$\epsilon_1(\omega) = 1 - \frac{\omega_p^2}{\omega(\omega + i/\tau)}. \tag{40}$$

In the perpendicular directions we assume a constant dielectric function

$$\epsilon_2 = \epsilon_3 = 1. \tag{41}$$

The resulting complex frequency-dependent $\epsilon_e(\omega)$, as given in the EMA, is shown in Fig. 1.

Fig. 2 shows the corresponding NDA prediction [eqs. (27)-(32)] for the enhancement of the cubic nonlinearity in the same polycrystal, under the assumption that $\epsilon_e$ (and $F(s)$) are given by the EMA. For comparison, we show in Fig. 3 the results of an earlier and less accurate approximation [4]. The results in Fig. 3 are obtained by using $|\langle E_i^2 \rangle|$ instead of $\langle |E_i|^2 \rangle$, and evaluating $\langle E_i^2 \rangle$ using eq. (29).

Quantitatively, based on our simple example, the effects of the correction noted by [3] appear to be relatively minor if only the diagonal elements $\chi_{iiii} \neq 0$, but are more substantial if $\chi_{1122} \neq 0$ or $\chi_{2211} \neq 0$. Nonetheless, the corrections do introduce a nonzero correction to all the elements of the tensor $\chi$. The numerical correction can be seen always to increase the absolute value of the corresponding matrix element of $\chi$. This trend can be qualitatively understood as follows: the spatial average $\langle E_i^2 \rangle$ which enters into the uncorrected matrix elements can, in principle, even vanish under some conditions, but the absolute value $\langle |E_i|^2 \rangle$ can never vanish.

## V. DISCUSSION

Next, we briefly discuss the limitations of the present approach, and the validity of the approximations made. At various points in this paper, we have made the following approximations:

- the quasistatic approximation;
- the nonlinear decoupling approximation; and
- the effective-medium approximation.

We now discuss the limitations of each of these approximations.

The *quasistatic approximation* is embodied in eq. (4), which implies that the electric field can be expressed as the negative gradient of a scalar potential. This assumption is still valid at finite frequencies, provided that the material of interest lies in the *long-wavelength limit* (see, for example, ref. [2]). A polycrystalline material is likely to fall in this long-wavelength regime, provided that the typical size of a crystallite is small compared to the wavelength of electromagnetic radiation in the medium. At optical frequencies, this condition requires crystallites of linear dimensions only a few hundred Å, but the same approximation might hold at microwave frequencies even for micron-size crystallites. More generally, any structural correlations in the polycrystal should exist only a scale which is small compared to the wavelength; otherwise, there is likely to be significant scattering of electromagnetic radiation and the quasistatic approximation will break down. Our derivation of the spectral representation for a polycrystal is based on the quasistatic approximation. Furthermore, our definition of an effective linear dielectric function $\epsilon_e$ [eq. (3)] presupposes the quasistatic approximation. Indeed, if the quasistatic approximation does not hold, the composite cannot easily be described in terms of an effective dielectric function.

The *nonlinear decoupling approximation* (NDA) is a way of approximately calculating the fourth moment of the electric field in the polycrystal, by breaking this up into a product of two second-moment terms. The NDA is known to be quite inaccurate near percolation thresholds in conventional composite materials and most likely also in polycrystalline materials [10,11]. The reason for the inaccuracy is that the NDA neglects local fluctuations in electric fields which become very important near a percolation threshold. However, if one does not make the NDA, than there is no easy way to express this nonlinear susceptibility in terms of the spectral function which describes the *linear* properties of the polycrystal.

Eqs. (27-30) and (32), which are based on our use of the quasistatic approximation and NDA, are equally valid for any microgeometry of a polycrystal and can be used with any desired approximation for the second moments. On the other hand, the spectral function $F(s)$, corresponding to the actual distribution of the electric field, may be sensitive to the particular arrangement of the crystallites. We use the *effective-medium approximation* (EMA) to calculate the spectral function which characterizes the linear dielectric function $\epsilon_e$. Although the EMA does predict the occurrence of a percolation threshold in a polycrystalline material, the approximation is likely to be quite inaccurate near that threshold, since it treats each crystallite as being embedded in an effective environment. Note that the spectral representation itself is more general than the EMA, since it is always applicable in the quasistatic approximation. Hence, the spectral representation can be used in conjunction with other, more accurate methods of calculating the linear response $\epsilon_e$, which take better account of local environment of a given crystallite, if such methods can be found.

We emphasize again that neither the NDA nor the EMA are necessary approximations; if better approximations for the fourth moment and for the linear response are available, then these can be used to compute the cubic nonlinear susceptibility of a polycrystalline material. The inaccuracy of the NDA and the EMA is partially compensated by the simplicity of these approximations, which allow many properties to be computed nearly in closed, analytic form.

To summarize, in this paper, we have extended the spectral representation of Bergman so that it applies to the linear effective dielectric function of a uniaxial polycrystal. The extension is straightforward, but should be useful in a wide variety of materials. As an illustration, we give the spectral function for a polycrystal in which $\epsilon_e$ is given in the effective-medium approximation. Finally, we use this spectral function to calculate the cubic nonlinear susceptibility tensor $\chi_e$ for a uniaxial polycrystal in the nonlinear decoupling approximation (NDA), once again calculating the required electric field averages within the EMA. As in two-component composites of isotropic materials, the expressions for $\chi_e$ are slightly altered from previous results when one properly accounts for the fact [3] that the averages $\langle E^2 \rangle$ and $\langle |E|^2 \rangle$ are unequal. We also give a brief discussion of the conditions under which these various expressions and approximations are applicable to real polycrystalline materials.

## VI. ACKNOWLEDGMENTS

This work was supported by the National Science Foundation, Grant DMR97-31511.

# REFERENCES

[1] D. J. Bergman Phys. Rep. **43**, 377 (1978); D. J. Bergman Phys. Rev. B **19**, 2359 (1979)

[2] D. J. Bergman and D. Stroud, in *Solid State Physics*, edited by H. Ehrenreich and D. Turnbull (Academic, New York, 1992), Vol. 46, pp. 178-320

[3] H. Ma *et al*, J. Opt. Soc. Am. B **15**, 1022 (1998)

[4] D. Stroud, Phys. Rev. B **54**, 3295 (1996)

[5] D. Stroud and P. M. Hui, Phys. Rev. B **37**, 8719 (1988)

[6] D. Stroud and A. Kazaryan, Phys. Rev. B **53**, 7076 (1996)

[7] A. V. Butenko *et al*, Z. Phys. D **17**, 283 (1990); M. I. Stockman *et al*, Phys. Rev. Lett. **72**, 2486 (1994)

[8] X. C. Zeng *et. al.*, Phys. Rev. B **38**, 10 970 (1988); X. C. Zeng *et. al.*, Physica A **157**, 192 (1989)

[9] D. A. G. Bruggeman, Ann. Phys. (Leipzig) **24**, 636 (1935); R. Landauer, J. Appl. Phys. **23**, 779 (1952)

[10] D. J. Bergman, Phys. Rev. **B39**, 4598 (1989)

[11] V. Shalaev and A. Sarychev, Phys. Rev. **B57**, 13265 (1998).

FIGURES

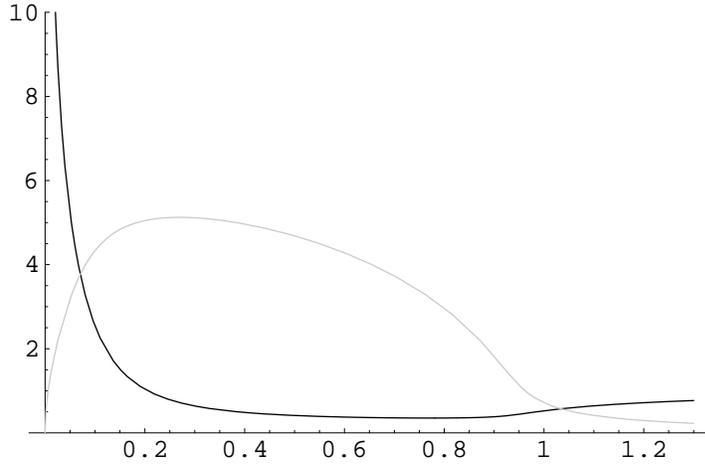

FIG. 1. $Re(\epsilon_e)$ (bold line) and $100 Re(\sigma_e)$ (light line), as given by EMA for a polycrystalline sample of a quasi-1D conductor. The single-crystal dielectric tensor is assumed to have principal values given by Eqs. (40) and (41) with $\omega_p\tau = 10$. $\sigma_e$ is defined by $\sigma_e \equiv \frac{i\omega}{4\pi}\epsilon_e$.

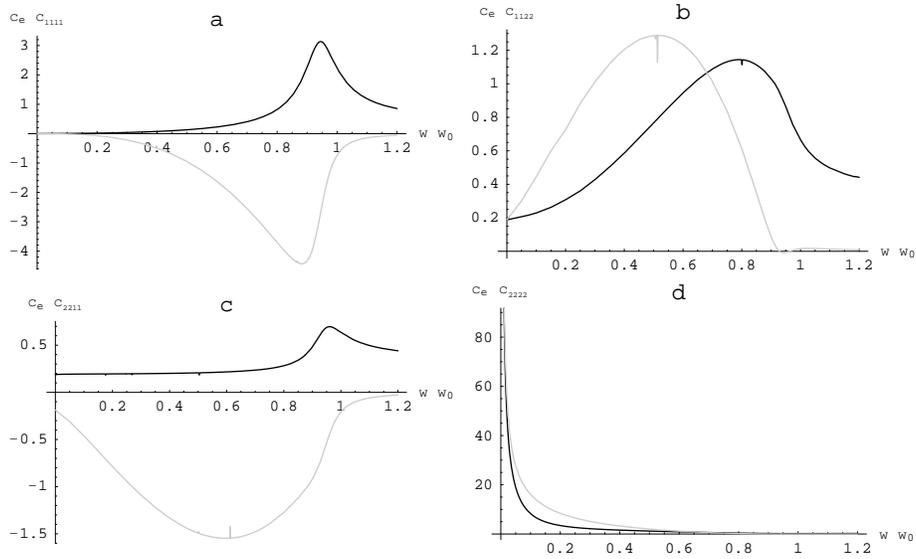

FIG. 2. (a) Real and imaginary parts of $\chi_e/\chi_{1111}$ (bold and light curves) for a polycrystalline material, calculated under the assumption that the only nonzero component of the single-crystal nonlinear susceptibility tensor is $\chi_{1111}$ (axis 1 parallel to the high-conductivity axis). The calculations are based on EMA results for the linear response (see Fig. 1), and on the model single-crystal dielectric tensor assumed in that Figure. (b) Same as (a) except that $\chi_e/\chi_{1122}$ is plotted, assuming that only $\chi_{1122}$ is nonzero. [The same plot will describe the enhancement of $\chi_{1212}$ and $\chi_{2112}$, as follows from the definition (23).] (c) Same as (a) except that $\chi_e/\chi_{2211}$ is plotted, assuming that only $\chi_{2211}$ is nonzero. (The same plot will describe the enhancement of $\chi_{1221}$ and $\chi_{2121}$.) (d) Same as (a) except that $\chi_e/\chi_{2222}$ is plotted, assuming that only $\chi_{2222}$ is nonzero.

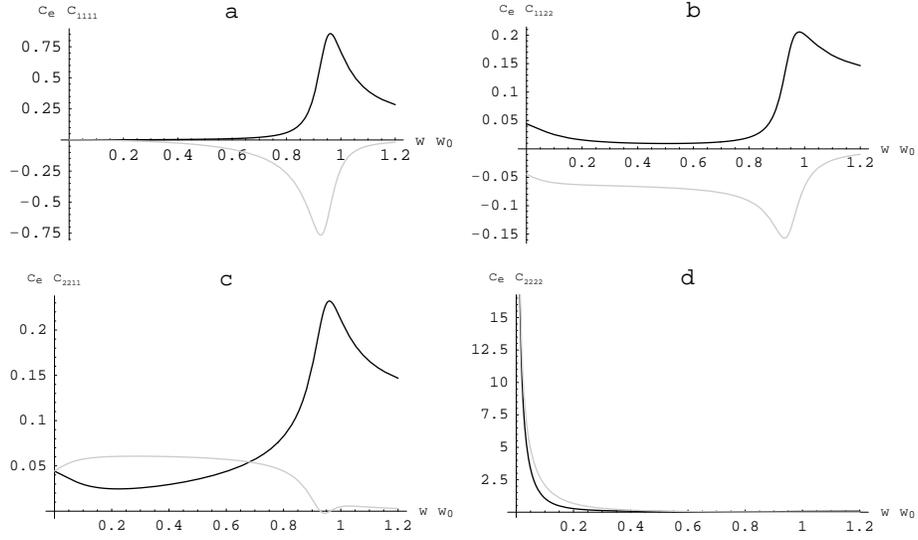

FIG. 3. Same as Fig. 2, except that instead of the more accurate formulas used in that Figure we use the expressions given in Ref. [4]. Plots (a)-(d) correspond to the plots (a)-(d) of Fig. 2.

14